\documentclass[twocolumn]{aastex62}

\graphicspath{{./}{figures/}}

\shorttitle{Hidden Planets: Implications from 'Oumuamua and DSHARP}
\shortauthors{Rice \& Laughlin}

\begin{document}

\title{Hidden Planets: Implications from 'Oumuamua and DSHARP}

\author[0000-0002-7670-670X]{Malena Rice}
\affiliation{Department of Astronomy, Yale University, New Haven, CT 06511, USA}
\affiliation{NSF Graduate Research Fellow}

\author{Gregory Laughlin}
\affiliation{Department of Astronomy, Yale University, New Haven, CT 06511, USA}

\correspondingauthor{Malena Rice}
\email{malena.rice@yale.edu}

\begin{abstract}
The discovery of 'Oumuamua (1I/2017 U1), the first interstellar interloper, suggests an abundance of free-floating small bodies whose ejection into galactic space cannot be explained by the current population of confirmed exoplanets. Shortly after 'Oumuamua's discovery, observational results from the DSHARP survey illustrated the near-ubiquity of ring/gap substructures within protoplanetary disks, strongly suggesting the existence of a vast population of as-yet undetected wide-separation planets that are capable of efficiently ejecting debris from their environments. These planets have $a \gtrsim 5$ au and masses of order Neptune's or larger, and they may accompany $\sim$50\% of newly formed stars \citep{zhang2018disk}. We combine the DSHARP results with statistical constraints from current time-domain surveys to quantify the population of detectable icy planetesimals ejected by disk-embedded giant planets through gravity assists. Assessment of the expected statistical distribution of interstellar objects is critical to accurately plan for and interpret future detections. We show that the number density of interstellar objects implied by 'Oumuamua is consistent with 'Oumuamua itself having originated as an icy planetesimal ejected from a DSHARP-type system via gravity assists, with the caveat that 'Oumuamua's lack of observed outgassing remains in strong tension with a cometary origin. Under this interpretation, 'Oumuamua's detection points towards a large number of long-period giant planets in extrasolar systems, supporting the hypothesis that the observed gaps in protoplanetary disks are carved by planets. In the case that 'Oumuamua is an ejected cometary planetesimal, we conclude that LSST should detect up to a few interstellar objects per year of 'Oumuamua's size or larger and over 100 yr$^{-1}$ for objects with $r > 1\,{\rm m}$.
\end{abstract}


\section{Introduction} \label{sec:intro}

In October 2017, 1I/2017 U1, now 'Oumuamua, was identified by the Panoramic Survey Telescope and Rapid Response System \citep[Pan-STARRS]{chambers2016} as the first interstellar interloper observed traversing the solar system \citep{meech2017}. 'Oumuamua's measured eccentricity $e=1.1956\pm0.0006$ places its trajectory firmly in the regime of hyperbolic orbits. It was on its way out of the solar system with heliocentric distance $d_\mathrm{H} = 1.22$ au upon its discovery, having passed perihelion with $d_\mathrm{H} = 0.25$ au at closest approach \citep{meech2017}. There is no clear consensus about the exact nature of 'Oumuamua, due in part to the short window of observability after its discovery and the corresponding sparsity of data, as well as the seemingly conflicting lines of evidence that have since emerged \citep{Sekanina2019}.

Following the detection of 'Oumuamua, several authors estimated occurrence rates for analogous interstellar objects (hereafter, ISOs) free-floating in galactic space, with all estimates falling within the range $n\approx0.004-0.24\,\mathrm{au}^{-3}$ deduced by \citet{portegies2018origin}). Assuming a cylindrical galaxy with $10^{11}$ stars, $R=3\times10^4\,\mathrm{pc}$ and $H=10^3\,\mathrm{pc}$ as in \citet{laughlin2017consequences}, even the most conservative of these estimates, $n\approx0.004\,\mathrm{au}^{-3}$, results in at minimum $0.36 M_{\oplus}$ of free-floating material per star, while the highest estimate $n\approx0.24\,\mathrm{au}^{-3}$ implies over $20 M_{\oplus}$ per star. Throughout this work, we adopt the fiducial estimate $n=0.2\,\mathrm{au}^{-3}$ from \citet{do2018interstellar}, which carefully incorporates the volume probed by Pan-STARRS to provide the most robust estimate to date.

'Oumuamua's detection appears to require a high density of material ejected into interstellar space, though both the ejection mechanism and the origins of the body remain under debate \citep[e.g.][]{bannister2019natural}. No coma, nor any cometary molecular emission bands, was detected in association with 'Oumuamua \citep{meech2017, trilling2018spitzer}. \citet{micheli2018non}, however, found that 'Oumuamua accelerated out of the solar system at a rate that could not be explained by gravitational forces alone, suggesting a cometary nature with acceleration induced by out-gassing.

Icy material is ejected from stellar systems during multiple evolutionary stages. Early on, close interactions between stars forming in open clusters can liberate icy planetesimals from their circumstellar disks \citep{hands2019fate}. Later, after giant planets have formed in the system, debris is ejected through dynamical interactions with these planets \citep[e.g.][]{barclay2017demographics, raymond2018implications}. Much of a star's enveloping Oort cloud is ultimately shed during the post-main-sequence stage of a star's lifetime \citep[e.g.][]{veras2011great, do2018interstellar, torres2019galactic}, releasing further volatile-rich material.

In this paper, we quantify the rate of icy planetesimal ejection induced by gravity assists from circumstellar planets, with the aim of setting expectations for future observations from the Large Synoptic Survey Telescope \citep[LSST;][]{abell2009lsst}. This exercise is critical to accurately assess the arrival prospects for future interstellar objects, each of which provides a uniquely valuable window into the nature of other planetary systems. 

To accomplish this, we study planetesimal ejection rates from the long-period giant planet population suggested by the Disk Substructures at High Angular Resolution Project \citep[DSHARP;][]{andrews2018disk}, which surveyed 20 nearby protoplanetary disks at high resolution using the Atacama Large Millimeter Array (ALMA). Considering the case in which 'Oumuamua stems from the population of ejected volatile-rich planetesimals, we show that gravity assists by long-period giant planets are capable of reproducing the number density of interstellar asteroids implied by the detection of 'Oumuamua. We use our results to predict the detection rate by LSST for ISOs produced through this channel. 

\citet{moro2018origin} previously used multi-component power law models to explore a wide range of possible ISO size distributions consistent with 'Oumuamua's appearance. Our estimates build on this previous work in several ways:

\begin{itemize}
    \item \citet{moro2018origin} adopted relatively low giant planet occurrence rates ($f_{\mathrm{pl}}$ = 0.2 for A-K2 stars and $f_{\mathrm{pl}}$ = 0.03 for K2-M stars) based on radial velocity measurements probing planets with semi-major axes $a\lesssim 3$ au. The DSHARP survey, however, suggests an abundance of giant planets ($f_{\mathrm{pl}}\sim$ 0.5) at wider separation $a \gtrsim 5$ au. We focus on these long-period planets, which are substantially more effective planetesimal ejectors than their shorter-period counterparts.

    \item Changes in the boundaries of a size distribution ($r_{\mathrm{min}}$ and $r_{\mathrm{max}}$) can dramatically alter the resulting fit. In \citet{moro2018origin}, these boundaries were set as constant values $r_{\mathrm{min}} = 1000$ km and $r_{\mathrm{min}} = 1\, \mu$m, motivated by solar system models. We include no assumptions for the value of $r_{\mathrm{min}}$ and leave $r_{\mathrm{max}}$ as a free parameter within our models. We thus avoid assumptions about the degree of similarity between the size distributions of solar system KBOs and free-floating ISOs, which may originate from a range of collisional histories that do not necessarily resemble that of the Kuiper belt.

    \item Lastly, \citet{moro2018origin} assumes that 100\% of solid circumstellar material around single stars and wide binaries is ejected. We instead perform N-body simulations of several representative systems from the DSHARP survey to deduce the expected mass of ejected material.
\end{itemize}

We present our simulation setup in Section \ref{section:methods} and subsequent results in Section \ref{section:results}, including our final range of possible power law solutions. We then discuss implications of our findings for future detections in Section \ref{section:lsst} and sources of uncertainty in Section \ref{section:uncertainty} prior to concluding.

\section{Methods}
\label{section:methods}

\subsection{Planets as Interstellar Comet Ejectors}
\label{planetsISOs}
Ejection of planetesimals is a natural outcome of close encounters with Jupiter, as observed in $N$-body simulations of early solar system evolution \citep[e.g.][]{o2006terrestrial, levison2009contamination}. Yet, not all planets are capable of efficiently ejecting material from their circumstellar systems.  To readily expel material through gravity assists, a planet must have Safronov number $\Theta\gtrsim1$, with $\Theta$ given by

\begin{equation}
    \Theta = \frac{v_{esc}^2}{v_{orb}^2} = \frac{M_p a}{R_p M_*}\, .
\label{eq:grav_f}
\end{equation}
Here, $M_p$, $R_p$, and $a$ are the mass, radius, and semimajor axis of a planet, respectively, while $M_*$ is the mass of the host star. Notably, hot Jupiters and super Earths cannot efficiently eject material from their systems. Long-period ($a\gtrsim5\,{\rm au}$) planets of Neptune's mass or greater are the most effective ejectors; these planets, however, lie in a region of parameter space that is heavily disfavored by the detection biases of the transit and radial velocity (RV) methods. Direct imaging results show that occurrence rates for giant planets with orbital separation $10-100$ au are low, at $9^{+5}_{-4}$\% between $5-13 M_{\rm Jup}$ \citep{nielsen2019gemini}, indicating that the primary planetary ejectors may be less massive than the current direct imaging detection limits and/or lie within $\sim10\,\mathrm{au}$ of the host star.

Recently, a candidate population of such planets was inferred by \citet{zhang2018disk} using results from DSHARP. Disk substructures were near-universally found in the DSHARP sample, where axisymmetric gaps and rings are most common. The DSHARP sample is biased towards relatively large disks around massive stars; however, gaps and rings at comparable radii have also been identified in samples more representative of the average protoplanetary disk population \citep[e.g.][]{long2018gaps}, suggesting that they may be prevalent among the underlying population. Although these substructures could be induced by a number of mechanisms, mounting evidence for planetary companions in similar systems \citep[e.g. PDS 70;][]{haffert2019two} favors the hypothesis that the substructures in the DSHARP sample are caused by planets. Recent kinematic detections also point towards the presence of planets at radial locations coincident with observed gaps in protoplanetary disks \citep{pinte2019kinematic}, further strengthening the case that some of the observed substructures are indeed induced by planets.

Consequently, the DSHARP sample suggests a $\sim50\%$ occurrence rate for giant planets with masses between $\sim M_{\mathrm{Nep}}$ and a few $M_{\mathrm{Jup}}$ orbiting their host stars at separation $a\gtrsim5$ au \citep{zhang2018disk}. This abundance of long-period giant planets is roughly in agreement with previous results from \citet{bryan2016statistics}, which combined Keck RV measurements with NIRC2 adaptive optics imaging to obtain an occurrence rate of $52\pm5\%$ for planets with $M=1-20 M_{\mathrm{Jup}}$ and $a=5-20$ au. An abundance of Neptune-mass planets is also supported by microlensing results, which find a peak in planet occurrence rates at planet-to-star mass ratio $q\sim10^{-4}$, corresponding to roughly 20$M_{\oplus}$, or 1.2 $M_{\mathrm{Nep}}$, for typical host star mass $\sim0.6M_{\odot}$ \citep{suzuki2016exoplanet}.

\subsection{Simulation Setup}
\label{section:simulationsetup}

To determine how this planet population connects to the population of  interstellar asteroids, we complete a detailed assessment of mass ejection rates in three DSHARP disk systems with clear radial gaps -- HD 143006, AS 209, and HD 163296 -- using the \texttt{REBOUND} orbital integrator \citep{rein2012rebound}. We adopt host star masses and radial planet positions from Table 3 of \citet{zhang2018disk}. For each predicted planet, \citet{zhang2018disk} reports several masses derived with varying assumptions for the disk dust size distribution and $\alpha$ viscosity parameter. Because these model parameters are not well-constrained, we randomly assign the mass of each planet from the range $[M_{\rm min}, M_{\rm max}]$, where $M_{\rm min}$ and $M_{\rm max}$ are the minimum and maximum derived masses, respectively. In accordance with results from \citet{zhang2018disk}, the HD 143006 and AS 209 systems each include 2 planets, while the HD 163296 system includes 3 planets.

\begin{figure}
    \centering
    \includegraphics[width=0.5\textwidth]{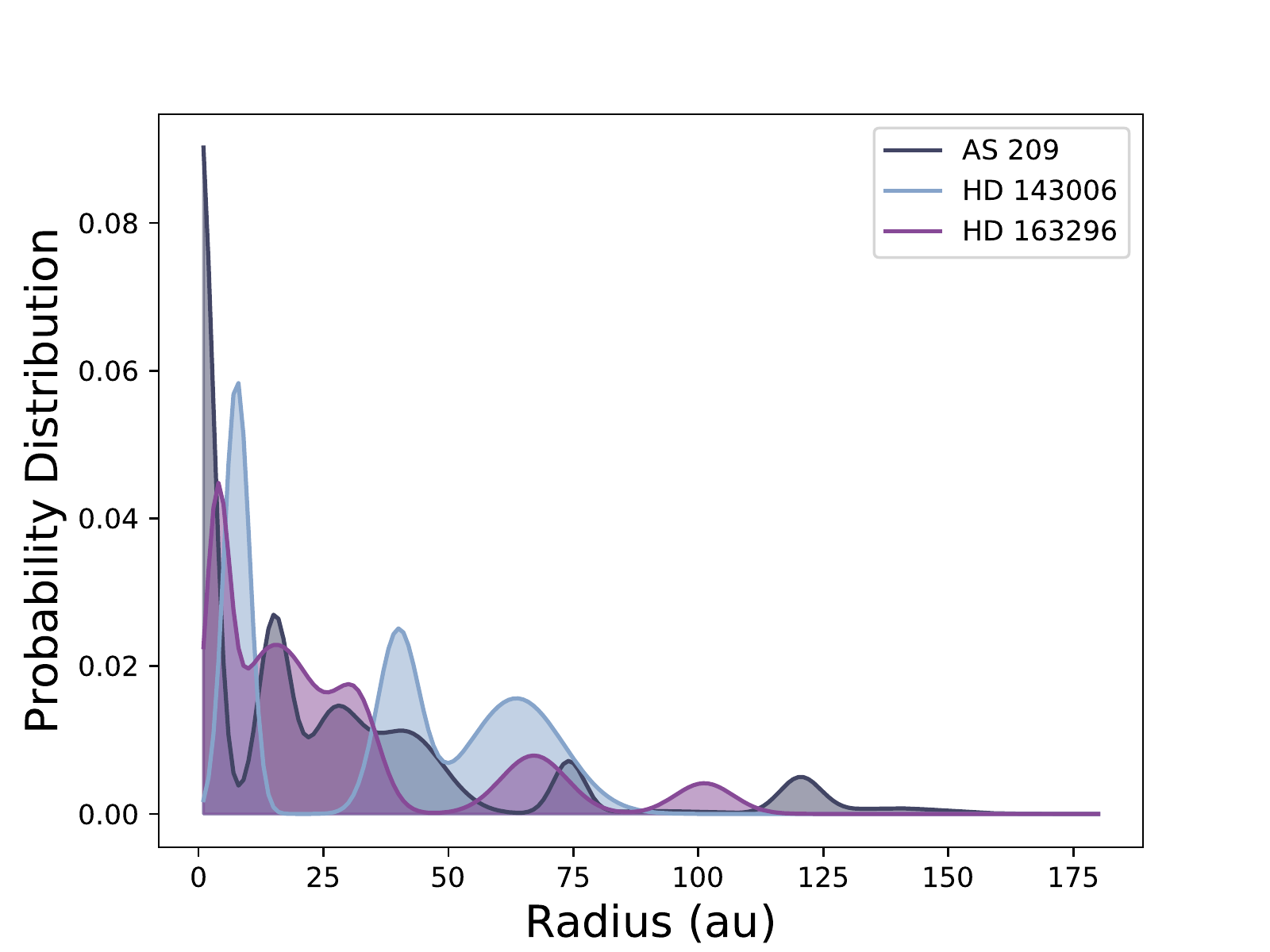}
    \caption{Initial distribution of dust particle semimajor axes for each disk in our sample. AS 209 is shown in gray, HD 143006 in blue, and HD 163296 in purple.}
    \label{fig:dust_distribution}
\end{figure}

We focus on the millimeter-sized dust population in our simulations, since dust masses in this size regime are relatively well-constrained by ALMA. We estimate the total mm dust mass $M_{\mathrm{dust}}$ of each disk using the relation \begin{equation}
    \mathrm{log}(M_{\mathrm{dust}}/M_{\oplus}) = 1.3\,\times\,\mathrm{log}(M_*/M_{\odot}) + 1.1
\label{eq:diskmass}
\end{equation} 
derived in \citet{pascucci2016steeper}, where $M_*$ is the mass of the host star. This derivation is based on the 887 $\mu$m flux measured from observations of the Chamaeleon I star-forming region, with bandwidth spanning $867-917\, \mu$m. Thus, the dust masses in our simulations represent only the subset of grains with radii in this size range.

To reproduce the dust mass distribution in each system, we initialize 3500 equal-mass test particles radially distributed using the semimajor axis probability distributions depicted in Figure \ref{fig:dust_distribution}. These distributions are based on radially symmetric best-fitting models consisting of several superposed Gaussians for each disk, with parameters for AS 209, HD 163296, and HD 143006 given in \citet{guzman2018disk}, \citet{isella2018disk}, and \citet{perez2018disk}, respectively. Each planetesimal is initialized with orbital elements

$$ \Omega=\omega=0$$
$$f\in[-\pi,\, +\pi]$$
$$i\in[0.0005,\, 0.005]$$
$$e\in[0.04,\, 0.06]$$
where all angles are in radians. We initialize each of our three disk-star-planet systems in three iterations each with a different random seed, resulting in a total of nine simulations with varying planet masses, orbital elements, and initial dust particle placements. We integrate each of these systems for one week on the \textit{Grace} supercomputing cluster at Yale University.

\section{Results}
\label{section:results}

\begin{figure}
    \centering
    \includegraphics[width=0.5\textwidth]{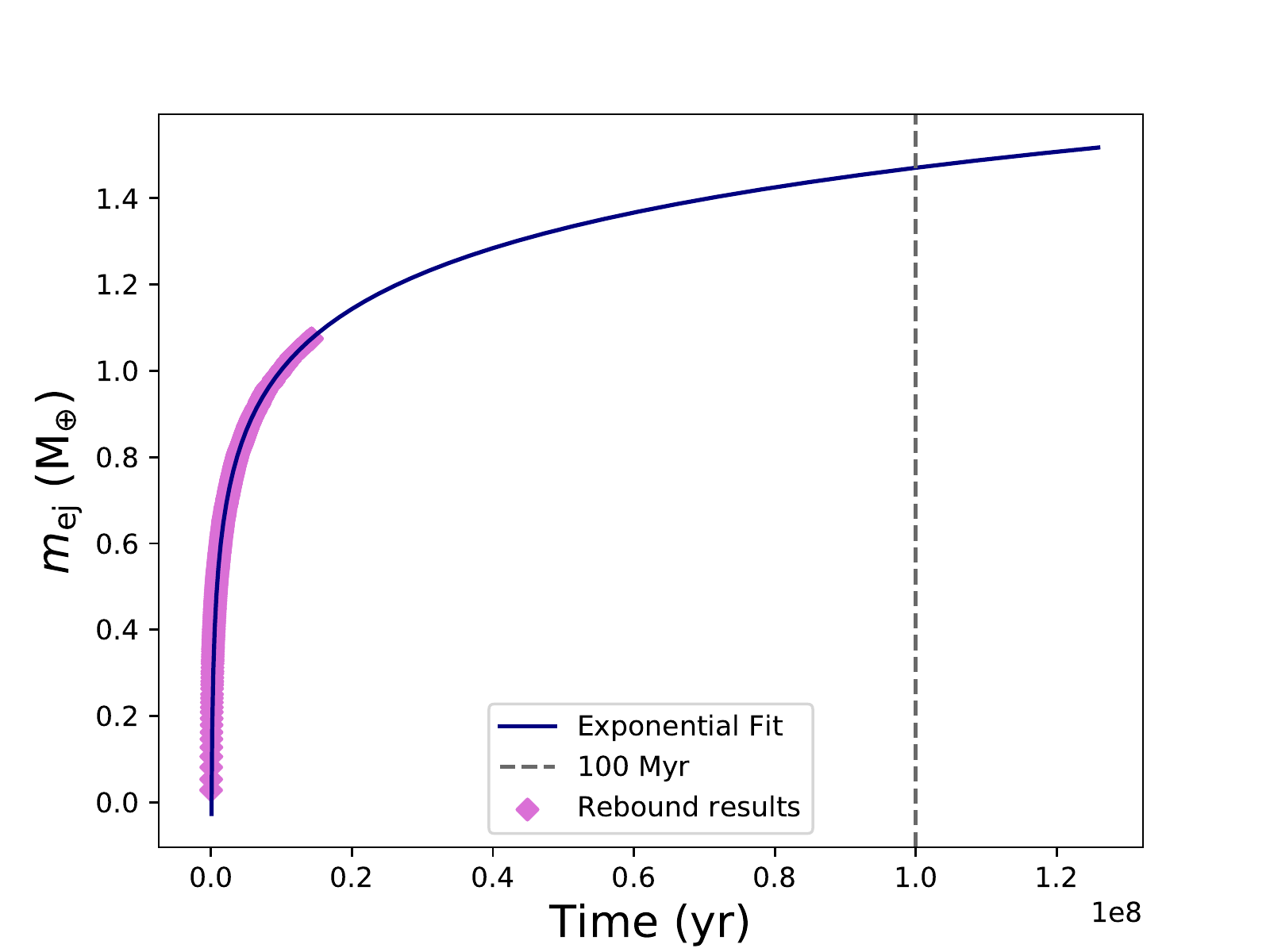}
    \caption{Average mass ejected as a function of time for all disks in our sample.}
    \label{fig:avg_m_fit}
\end{figure}

To encapsulate the typical behavior of all systems, we find the average mm mass $m_{\mathrm{ej}}$ ejected by each star as a function of time $t$ for all nine simulations, with our results displayed in Figure \ref{fig:avg_m_fit}. The resulting curve, representing the total mass of all particles with a formally positive energy relative to the host star, is well-fit by exponential function

\begin{equation}
    m_{\mathrm{ej}} = a\log_{10} t + b,
\label{eq:logmassprofile}
\end{equation}
with $a=0.468$ and $b=-2.274$. Using this profile, we extend our results to $t=10^8\,\mathrm{yr}$, at which point $m_{\mathrm{ej}}=1.47 M_{\oplus}$. We take this as the representative mm-sized dust mass ejected by each giant-planet-hosting star throughout its main-sequence lifetime, selecting a relatively late time $t$ to compensate for mass ejected prior to the gas-clearing phase in each system. Adopting a $50\%$ occurrence rate for this population of planets, we approximate that the average rate of mm-sized planetesimal ejection is $0.74\, M_{\oplus}\,\mathrm{star^{-1}}$.

In order to connect the mass ejected per star to an ISO spatial number density $n$, we must first determine the background stellar number density $n_*$. We adopt the local midplane stellar mass density $\rho_* = 0.043\pm0.004 \,M_{\odot}\, \mathrm{pc}^{-3}$ found in \citet{mckee2015stars}, then estimate the average stellar mass in the solar neighborhood using the sample of 75 known stars (including brown dwarfs) within 5 pc of the Sun. For stars without measured masses, we apply the piecewise mass-luminosity relation from \citet{eker2015main}. We obtain an average stellar mass $M \sim 0.35 M_{\odot}$, which translates to stellar number density $n_*\approx0.12\,\,\mathrm{stars}\,\,\mathrm{pc}^{-3}$. 

Combining $n_*$ with our average planetesimal mass ejection rate per star, we find projected ISO mass density $\rho_{\rm ISO} \sim0.09\,M_{\oplus}\,\mathrm{pc}^{-3}$ free-floating in space. This means that, for number density distribution $dn/dr$ with boundaries $r_{\rm min, obs} = 867\,\mu$m and $r_{\rm max, obs} = 917\,\mu$m,

\begin{equation}
    \int^{r_{\rm max, obs}}_{r_{\rm min, obs}} \frac{dn}{dr} m(r) dr = 0.09\,M_{\oplus}\,\mathrm{pc}^{-3}.
\label{eq:constraint_mass}
\end{equation}

We use $m(r)=\frac{4}{3}\pi r^3\rho$ with $\rho=1\,\mathrm{g}\,\mathrm{cm}^{-3}$ and fit $dn/dr$ with a power law radius distribution of the form

\begin{equation}
\frac{dn}{dr} = Cr^{-q},
\label{eq:powerlaw}
\end{equation}
where $q$ is the power law index and $C$ is constant. We interpret previous estimates of $n$ obtained from the detection rate implied by 'Oumuamua as $n(r\gtrsim r_{\rm O})$ where $r_{\rm O}$ is 'Oumuamua's effective radius, since larger ISOs would also be detectable when passing through the solar system at 'Oumuamua's distance. Equivalently,

\begin{equation}
\int^{r_\mathrm{max}}_{r_{\mathrm{O}}} \frac{dn}{dr}dr = n.
\label{eq:constraint_n}
\end{equation}

Combining and integrating Equations \ref{eq:constraint_mass}, \ref{eq:powerlaw}, and \ref{eq:constraint_n}, we parameterize our solution space with the function

$$ f = \frac{4}{3}\pi\rho n \Big( \frac{1}{0.09 M_{\oplus}\,\mathrm{pc}^{-3}} \Big) \Big(\frac{1-q}{4-q}\Big) \Big[\, \frac{r_{\rm max, obs}^{4-q} - r_{\rm min, obs}^{4-q}}{r_{\rm max}^{1-q} - r_{\rm O}^{1-q}}\,\Big], $$
which is equal to unity when Equations \ref{eq:constraint_mass}, \ref{eq:powerlaw}, and \ref{eq:constraint_n} are simultaneously satisfied. Using $f$, we solve for power law indices $q$ over a range of $r_{\rm max}$, with $r_{\rm O}=55\,\mathrm{m}$ from \citet{jewitt2017interstellar}.

\begin{figure}
    \centering
    \includegraphics[width=0.5\textwidth]{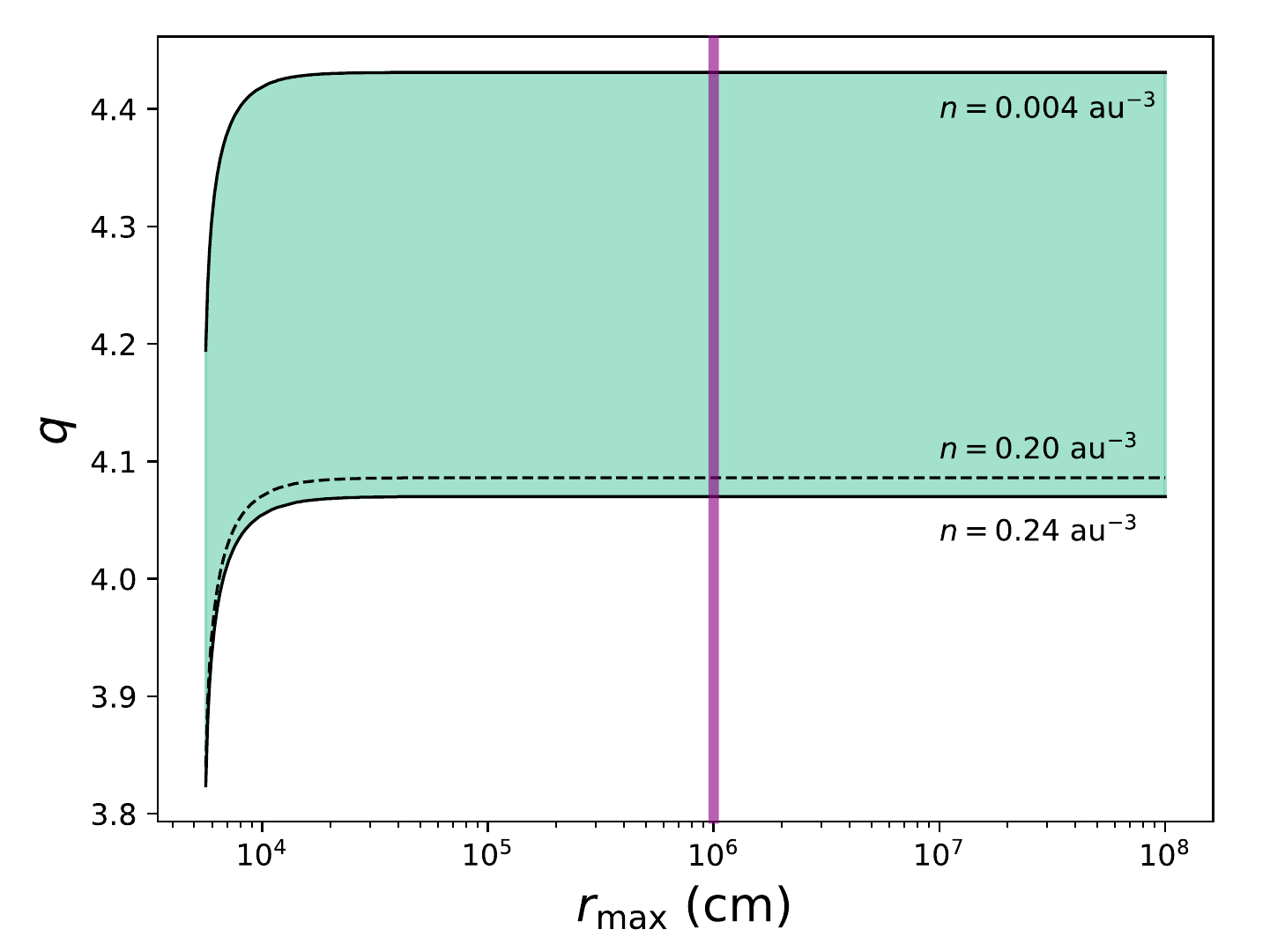}
    \caption{Solutions to the power law index $q$ as a function of $r_{\rm max}$ for the full range of $n$ estimates given in \citet{portegies2018origin}. The \citet{do2018interstellar} estimate $n=0.2$ au$^{-3}$ is indicated with a dotted line. The purple slice of solutions at $r_{\rm max}=10^6$ cm displays the cross-section corresponding to Figure \ref{fig:q_given_rmax}.}
    \label{fig:qmap_nrange}
\end{figure}

\begin{figure}
    \includegraphics[width=0.5\textwidth]{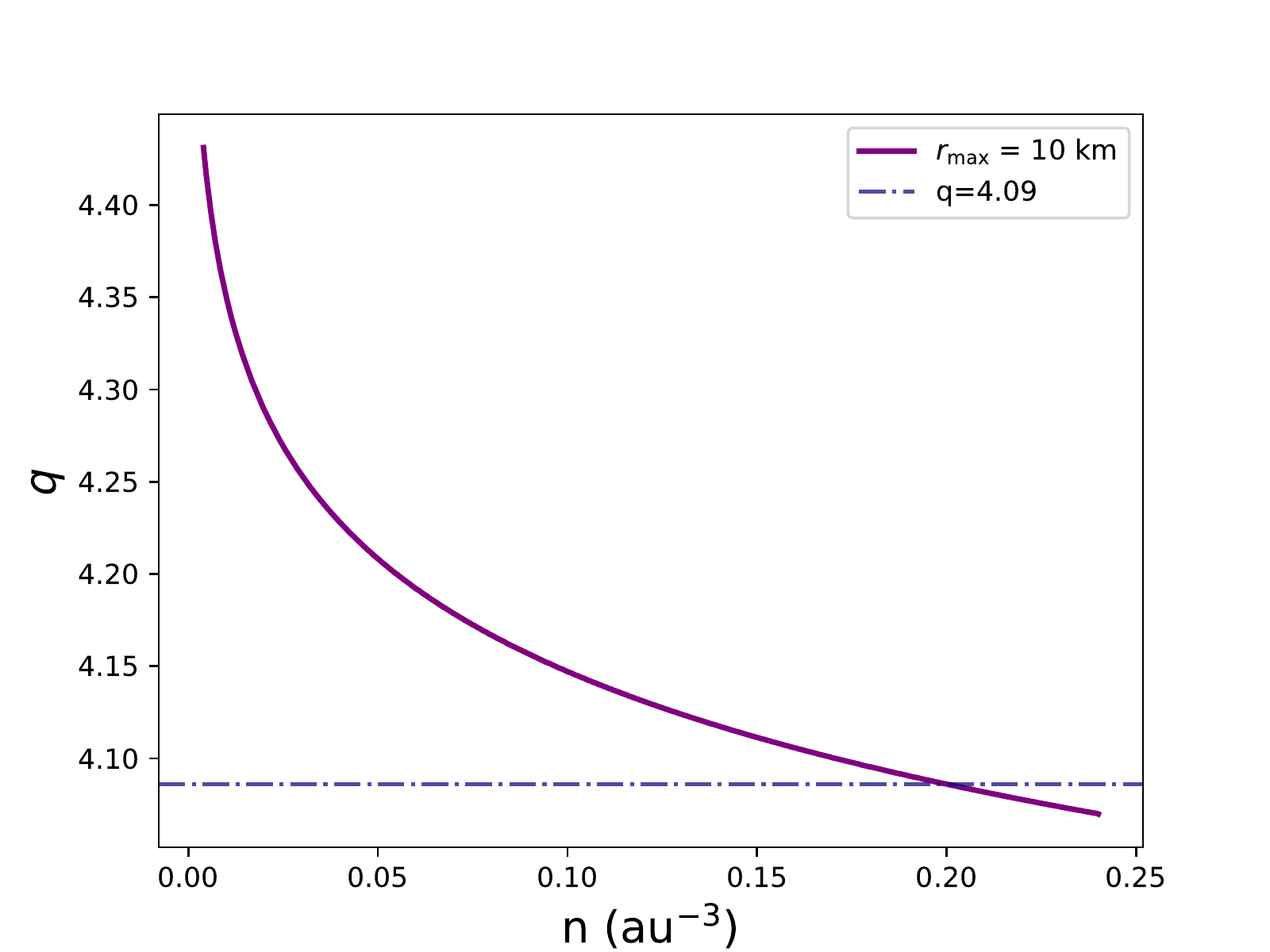}
    \caption{Power law solutions $q$ shown as a function of $n$. Here, we set $r_{\rm max}=10 \,\mathrm{km}$, noting that $q$ varies little with changes in $r_{\rm max}$ (see Figure \ref{fig:qmap_nrange}). The dot-dashed blue line denotes the solution where $n=0.2\,\mathrm{au}^{-3}$.}
    \label{fig:q_given_rmax}
\end{figure}

The resulting solutions are displayed in Figure \ref{fig:qmap_nrange}, indicating that, for a given $n$, our power law solutions are robust to changes in $r_{\mathrm{max}}$ for an ISO distribution in which $r_{\mathrm{max}}\not\approx r_{O}$. We take a vertical slice through Figure \ref{fig:qmap_nrange} at $r_{\mathrm{max}} = 10^6\,\mathrm{cm}$ to display $q$ as a function of $n$ in Figure \ref{fig:q_given_rmax}. Ultimately, we find that high power law indices $q>4$ are required to fit both the $n$ values implied by 'Oumuamua and our simulated mass ejection rates. Adopting $q=4.09$, this distribution corresponds to a total mass of ejected ISO material $\sim24M_{\oplus}$ per DSHARP-type system for ISOs in the size range $10^{-3}\,\mathrm{cm} \leq r \leq 10^6\,\mathrm{cm}$.

Another potential constraint for this distribution is the number density $n_{\mathrm{CNEOS}}=10^{6^{+0.75}_{-1.5}} \,\mathrm{au}^{-3}$ of $r_{\mathrm{CNEOS}}\gtrsim0.45$ m interstellar objects implied by the candidate interstellar meteor identified from the Center for Near-Earth Object Studies (CNEOS) bolide catalog in \citet{siraj2019discovery}. We note that \citet{devillepoix2018observation} reported that the United States government sensors used to discover this object are generally unreliable for orbit calculations. We include the candidate CNEOS meteor for completeness and primarily focus on determining whether the object is consistent with our analysis up to this point. If the CNEOS meteor originates from the same underlying distribution as 'Oumuamua, it would require

\begin{equation}
    \int^{r_{\rm max}}_{r_{\rm CNEOS}} \frac{dn}{dr}dr = n_{\mathrm{CNEOS}}.
\label{eq:CNEOS}
\end{equation}

Adopting $q=4.09$, corresponding to $n=0.2\,\mathrm{au^{-3}}$ from \citet{do2018interstellar}, we obtain number density $n=5.63\times 10^5 \,\mathrm{au}^{-3}$ for objects with $r_{\mathrm{CNEOS}} \leq r \leq r_{\mathrm{max}}$. This value is firmly within the error bars for $n_{\mathrm{CNEOS}}$ from \citet{siraj2019discovery}, indicating that our results are consistent with those implied by the candidate CNEOS meteor.

\section{Observability with LSST} 
\label{section:lsst}

Although the first ISO was not confirmed until 2017, anticipative estimates for the detectability of such objects have been made on several earlier occasions. To account for the prior non-detection of ISOs, early assessments predicted an interstellar comet number density $n\sim10^{13}$ pc$^{-3}$ for objects with $r\geq1$ km \citep{mcglynn1989nondetection}. Later, \citet{jewitt2003project} projected that, if all stars eject $10^{13}$ comets with $r\geq1\, \mathrm{km}$, approximately 0.3 such interstellar interlopers should pass within 5 au of the sun -- the approximate detection limit of Pan-STARRS -- each year.

\citet{moro2009will} followed up this study by calculating the expected detection rate by LSST of inactive extrasolar comets passing through the solar system at distances greater than 5 au. They found that the probability of LSST detecting an inactive interloper during its 10-year lifetime is $\sim0.01\%-1\%$, and they ultimately concluded that LSST will likely not observe even one such object. \citet{cook2016realistic} then updated prior estimates by numerically simulating the detectability of interstellar asteroids within 5 au of the sun, concluding that an optimistic estimate would result in 1 interstellar object detected during LSST's lifetime. Since 'Oumuamua's discovery with Pan-STARRS, \citet{trilling2017} used a scaling argument to project that the detection rate of analogous interlopers will increase from 0.2 yr$^{-1}$ with Pan-STARRS to 1 yr$^{-1}$ with the advent of LSST.

Leveraging the power law radius distributions obtained in Section \ref{section:results}, we independently estimate the expected detection rate by LSST for protostellar disk-ejected ISOs. Adopting single-frame magnitude limit $m_{\mathrm{LSST}} \sim 24$ \citep{lsst2009} and cometary geometric albedo $p_V=0.04$ \citep{lamy2004}, we calculate the smallest ISO radius observable with LSST at a given distance $d$, given in au, using

\begin{equation}
    r_{\mathrm{min, LSST}} = \frac{1\,\mathrm{au}}{\sqrt{p_v}}\,10^{\frac{1}{5}(m_{\odot} - H)}
\label{eq:detectability_limit}
\end{equation}

\begin{equation}
H = m_{\mathrm{LSST}} - 5\log_{10}d,
\end{equation}
where $m_{\odot}=-26.74$ is the magnitude of the Sun and $H$ is the absolute magnitude of the ISO. Our calculation implicitly assumes that all ISOs are observed at opposition; however, in practice the solar angle and frequency of sky coverage will also play an important role in the final ISO detection rate, meaning that our rates should be treated as upper limits.

\begin{figure}
    \centering
    \includegraphics[width=0.48\textwidth]{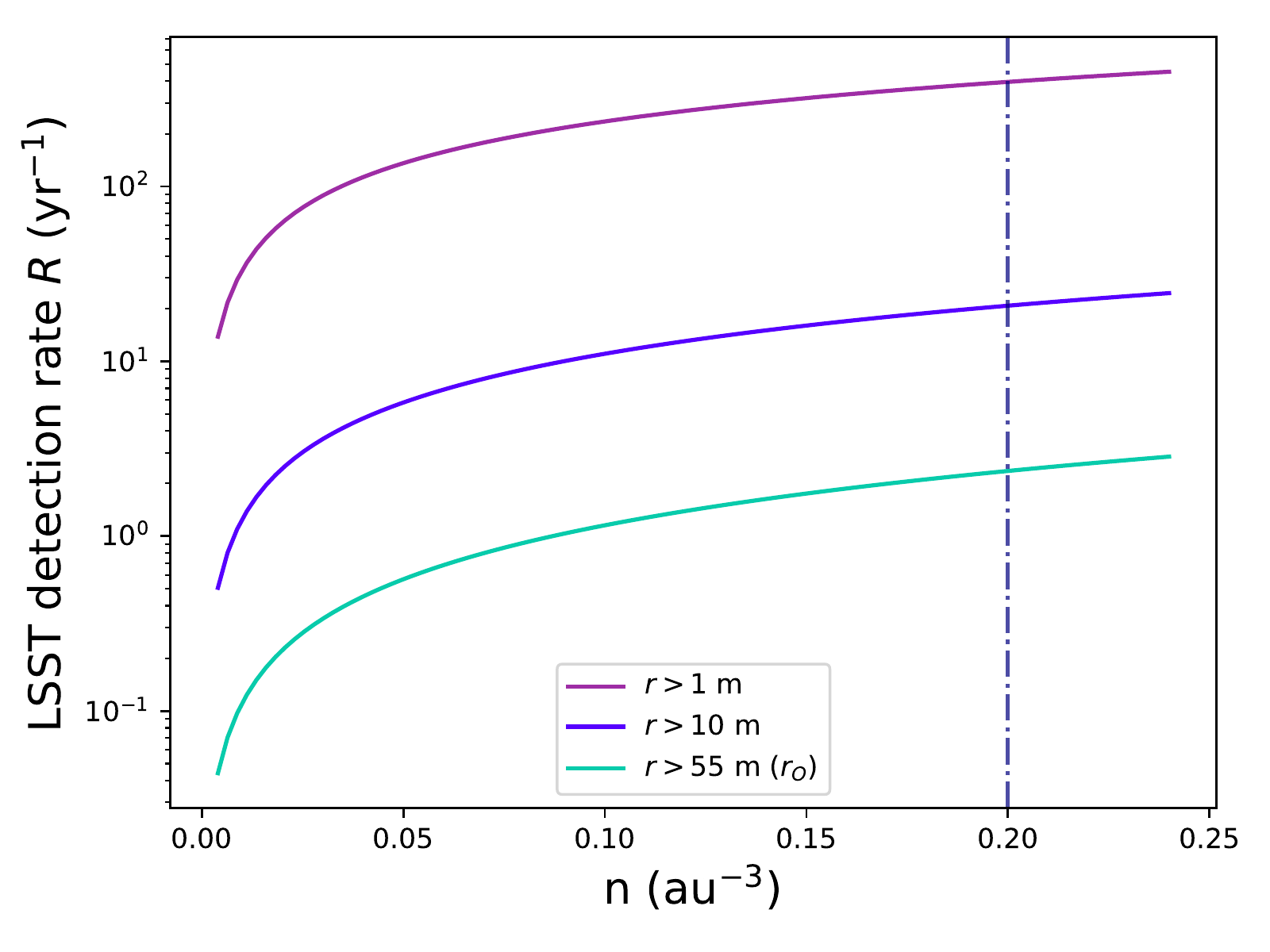}
    \caption{LSST detection rates as a function of $n$ for minimum object radii 1 m, 10 m, and $r_O=55 \,\mathrm{m}$.}
    \label{fig:lsst_rates}
\end{figure}

At each of $10^5$ evenly spaced circular annuli around the Earth, from $0-40$ au, we sum over the number density profile from $r_{\mathrm{min, LSST}}$ to $r_{\mathrm{max}}$ to obtain the total observable ISO number density $n_{\mathrm{ISO}}$ within that annulus. Combining $n_{\mathrm{ISO}}$ with the surface area of each annulus $\sigma$, we calculate the total detection rate of ISOs as

\begin{equation}
    R=\sum n_{\mathrm{ISO}}\sigma v,
\end{equation}
summing over all annuli. We take 'Oumuamua's velocity at infinity, 26 km/s \citep{mamajek2017}, as our representative $v$. Completing this calculation with reference minimum radius thresholds $r>1\,\mathrm{m}$, $r>10\,\mathrm{m}$, and $r>55\,\mathrm{m}$, we obtain the results outlined in Figure \ref{fig:lsst_rates}.

We find that, if 'Oumuamua is an ejected cometary planetesimal drawn from an isotropic power law distribution, LSST will find several ISOs each year of 'Oumuamua's size or larger, as well as up to hundreds of ISOs per year with $r>1\,\mathrm{m}$.

\section{Sources of Uncertainty}
\label{section:uncertainty}
\citet{zhu2019one} recently found that protoplanetary disk masses may be heavily underestimated due to the assumption that the disks are optically thin. As a result, in the case that these disks are optically thick, our mm mass ejection rates may be underestimated, leading to an underestimate in the power law slope.

However, it may instead be the case that our disk masses are overestimated. Our three sample disks are located around relatively massive stars ($0.83M_{\odot}, 1.78M_{\odot},$ and $2.04M_{\odot}$ for AS 209, HD 143006, and HD 163296, respectively), meaning that treating these systems as `typical' may instead lead to an overestimate of the total mass of ejected material. This would accordingly correspond to an overestimate of the LSST detection rate, particularly on the lower end of the ISO size distribution.

Furthermore, the number density of ISOs, $n$, may be locally enhanced at the present time due to the Sun's current proximity to the Galactic midplane: $z_{\odot}\sim 17 \pm 5$ pc \citep{karim2016revised}, as compared to its maximum height above the midplane $49-93$ pc \citep{bahcall1985sun}. If the Sun is currently passing through a low-dispersion, `cold' population of ISOs, the apparent number density as observed from the solar system may be higher than the bulk average density throughout the Galaxy. Because we do not yet have strong constraints on the scale height of ISOs, we cannot rule out this possibility, which suggests that the true $n$ may be lower than most current estimates. This could result in a steeper power law size distribution and a correspondingly lower LSST detection rate for large ISOs.

It is also quite possible that the distribution of interstellar comets does not follow a simple power law. \citet{moro2018origin}, after sampling a wide range of possible models, finds that 'Oumuamua is likely not representative of an isotropic background population. Though our present model is already overconstrained and therefore does not warrant the inclusion of additional free parameters, we acknowledge that a power law fit may not fully capture the true characteristics of the ejected ISO population. We may be overestimating the number density of large ISOs if, for example, the collisional evolution of solid material in extrasolar systems typically grinds the vast majority of larger bodies into dust prior to ejection.

Lastly, the detection rates reported in Section \ref{section:lsst} may be modestly increased by gravitational focusing and comet brightening from passage near the sun, each of which enhances observability of the ISO population.

\section{Conclusions} 
\label{section:conclusions}

The discovery of the first interstellar interloper, 'Oumuamua, has already provided exceptional insight to inform our understanding of planetary systems while simultaneously presenting new puzzles. In our work, we have reconciled simulations of the observed DSHARP planet-disk systems with the detection of 'Oumuamua to constrain the range of possible size distributions for interstellar objects ejected through interactions with circumstellar giant planets. We conclude that the population of long-period giant planets suggested by the DSHARP sample is capable of ejecting the population of free-floating planetesimals implied by 'Oumuamua's detection. Furthermore, 'Oumuamua is consistent with an origin within a population of ISOs following a single power law radius distribution, resulting in an anticipated LSST detection rate ranging from just a few detections per year for 'Oumuamua-sized ISOs to over $100\,\mathrm{yr}^{-1}$ for ISOs with $r > 1$ m. Future observations of interstellar objects hold tremendous potential to answer long-standing questions about not only the range of processes taking place in extrasolar systems, but also the population statistics of long-period giant exoplanets, the diversity of small-body populations throughout the Galaxy, and the evolutionary path of the solar system itself.

\section{Acknowledgements}
\label{section:acknowledgements}

We thank the anonymous referee for helpful comments that substantially improved the manuscript. We also thank Matt Holman for insightful input as the manuscript was being finalized. M.R. is supported by the National Science Foundation Graduate Research Fellowship Program under Grant Number DGE-1752134. This material is also based upon work supported by the National Aeronautics and Space Administration through the NASA Astrobiology Institute under Cooperative Agreement Notice NNH13ZDA017C issued through the Science Mission Directorate. We acknowledge support from the NASA Astrobiology Institute through a cooperative agreement between NASA Ames Research Center and Yale University. Simulations in this paper made use of the \texttt{REBOUND} code which can be downloaded freely at \url{http://github.com/hannorein/rebound}. This research also made use of the \texttt{numpy} \citep{oliphant2006guide}, \texttt{matplotlib} \citep{hunter2007matplotlib}, and \texttt{seaborn} \citep{waskom2014seaborn} Python packages.

\bibliography{bibliography}
\bibliographystyle{aasjournal}

\end{document}